# The Fragility of the Reverse Facilitation Effect in the Stroop Task: A Dynamic Neurocognitive Model

**Ahmad Sohrabi**

**Robert L. West**

**Carleton University, Ottawa, ON, Canada**

## Abstract

In typical Stroop experiments, participants perform better when the colors and words are congruent compared to when they are incongruent or neutral. Paradoxically, in some experimental conditions, neutral trials are faster than congruent trials. This phenomenon is known as Reverse Facilitation Effect (RFE). However, RFE has not been consistently replicated, leading to the so-called 'fragile' results. There are some models that capture this effect, but they are not parsimonious. Here we employed our previous model of priming effect, without its conflict monitoring module, to demonstrate that RFE, including its fragility, is mainly due to attentional dynamics with limited resources. The simulations of the RFE resulted from weak attentional activation for neutral stimuli (e.g., non-color or nonword strings) that causes less attentional refraction for the target stimuli (e.g., font colors). The refractory period or attenuation of attention usually happens when meaningful stimuli (color name primes) are used along font color targets. This simpler model provides a straightforward understanding of the underlying processes.

## Introduction

Human cognition operates through fast and slow processes, with their involvements or division of labor varying based on the stimuli and task condition. The dominance of each process is also known as System 1, which is fast and automatic, and System 2, which is slower and more controlled (e.g., Conway-Smith and West, 2022). We assume that both systems are usually active, but System 2 mobilization is driven by conflicting stimuli, features, or task demands. In the present study we aimed to focus on System 1 and exclude conflict or interference for the sake of simplicity to shed light on the dynamism and fluctuations in this domain.

For this purpose, we explore the Stroop task (Stroop, 1935; see also MacLeod, 1991), where recognizing the color of a congruent color word (e.g., **GREEN** in green font) is easier than recognizing the color of an incongruent color word (e.g., **RED** in green font). The resulting difference in performance, especially speed (i.e., RTs), is known as the incongruency (interference) effect. This effect can be taken as an index of system domination, something that is highly altered in mental disorders such as schizophrenia (e.g., Servan-Shreiber et al., 1990). However, the performance in congruent condition is not usually better than the neutral condition, especially when colored words are non-color names (e.g., word **BOOK** in green font) instead of colored non-words, random strings, or repeated characters (e.g., **RILT**, **NRPT**, or **XXX** in green font). So, paradoxically the speed in the neutral trials is faster than the congruent ones and is known as Reverse Facilitation Effect (RFE). Goldfarb and Henik (2007) revealed this effect simply by increasing the proportions of neutral stimuli among congruent or incongruent color words. But, the RFE has not been repeated consistently. Kalanthroff and Henik (2013) attributed the 'fragility' or inconsistent results to individual differences. In their recent model, Kalanthroff et al. (2018) simulated RFE by referring to Kornblum and colleagues' idea of different kinds of conflict (Kornblum, 1994; Kornblum and Lee, 1995). In the current study, we employ our previous model of priming effects (e.g., Sohrabi and West, 2009) without conflict monitoring module, to demonstrate that RFE including its fragility is mainly due to attentional dynamics and limited processing resources.

Several studies support the dynamic and multi-level view (Ritz et al., 2021; Verbeke and Verguts, 2020; Markovic et al., 2021). Here, to uncover the main underlying processes in RFE, we focus on intra-trial attention, presumably having minimum conflict, compared to planned and context-dependent tasks (Ott et al., 2021) where System 2 is likely more involved. Fluctuations in attention are mainly driven by System 1, especially the dynamic thalamocortical processes and neuromodulatory activities of Locus Coeruleus (LC), though still under investigations (Dahl et al., 2021). Therefore, another aim of this study was to meticulously explore the attentional dynamics through a simple model, detached from higher-level controls including conflict monitoring. Getting the model to work without conflict monitoring can provide a better understanding of the underlying processes, by avoiding irrelevant complexities. This can lead to a more basic model but generalizable to a broad range of cognitive functions that have been usually experimented and modelled separately.

It is hypothesized that a neutral stimulus, carrying less meaning, has no obvious interfering/facilitating effects on a target stimulus. However, being less strong, a neutral stimulus (prime) invokes less attentional resources, allowing more resources for the second stimulus (target). We are looking at how the interplay between the activations is related to the two stimuli that drive and are driven by the resulting attentional fluctuation, according to their overall strength of representation.

# Methods

## The Architecture of the Model

The architecture of the model is shown in Figure 1. The model consists of an Input Layer (IL), a Representation Layer (RL), an Attentional Layer (AL), and finally a Motor Layer (ML). The neuronal units in all main layers (except IL, static sensory) are Leaky Integrate and Fire (LIF). They simulate neural feed-forward activation, self-excitation, i.e., recurrent feedback, and lateral inhibition (not shown in the figure). However, the Attention Layer (AL), being a Wilson-Covan's (1972) dynamic system, is an abstract neural layer and provides phasic attentional activation (representing LC region of the brain). While RL units directly and selectively contribute to ML units, and hence a response, the AL works through neuromodulation, a sort of multiplicative boosting, influencing all layers, except IL. All layers are activated together dynamically but there are also thresholds that can prevent irrelevant and premature activations, as will be discussed and illustrated.

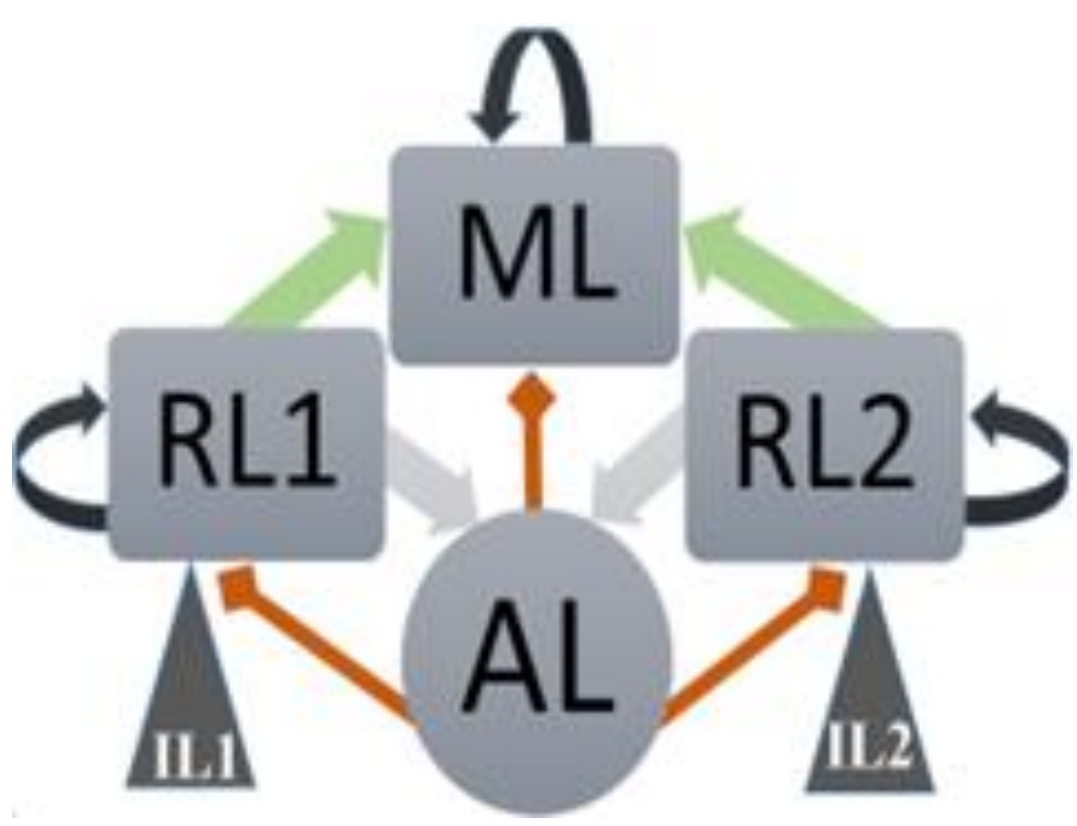

Figure 1: The architecture of the model shows hypothetical networks and connections. Layer types: Input Layers (ILs), shown as triangles (IL1 and IL2 for two stimuli or features), Representation Layer (RL) and Motor Layer (ML), shown as rectangles, and Attention Layer (AL), shown as a circle as well as connections including feedforward (straight arrows), self-excitation (curved arrows), and modulation (diamond-head arrows).

## Strength Representation

As shown in Figure 1, the manipulations of connection weights (straight arrows) are intended as a way of manipulating the strength of representation. This affects the dynamic activation of the relevant units as depicted in Figure 2 and 3, for congruent and incongruent trials respectively. Moreover, the weaker the stimulus, the less meaningful it is, and vice versa. So, the weakest can represent neutral stimuli, i.e., the word BED (with red fonts) and the strongest as a color word in the congruent trials, the word RED (with red font color). The difference between congruent and incongruent trials decreases with the level of strength or meaning (shown from left to right of each row in the figures). The target (font color with strength 3.0) is preceded by words with different levels of meaning or strength (1.5, 2.0, 2.5, and 3.0), from left to right, respectively. The activation of neutral prime should disappear when it becomes completely neutral (e.g., strengths less than 1.5, not used).

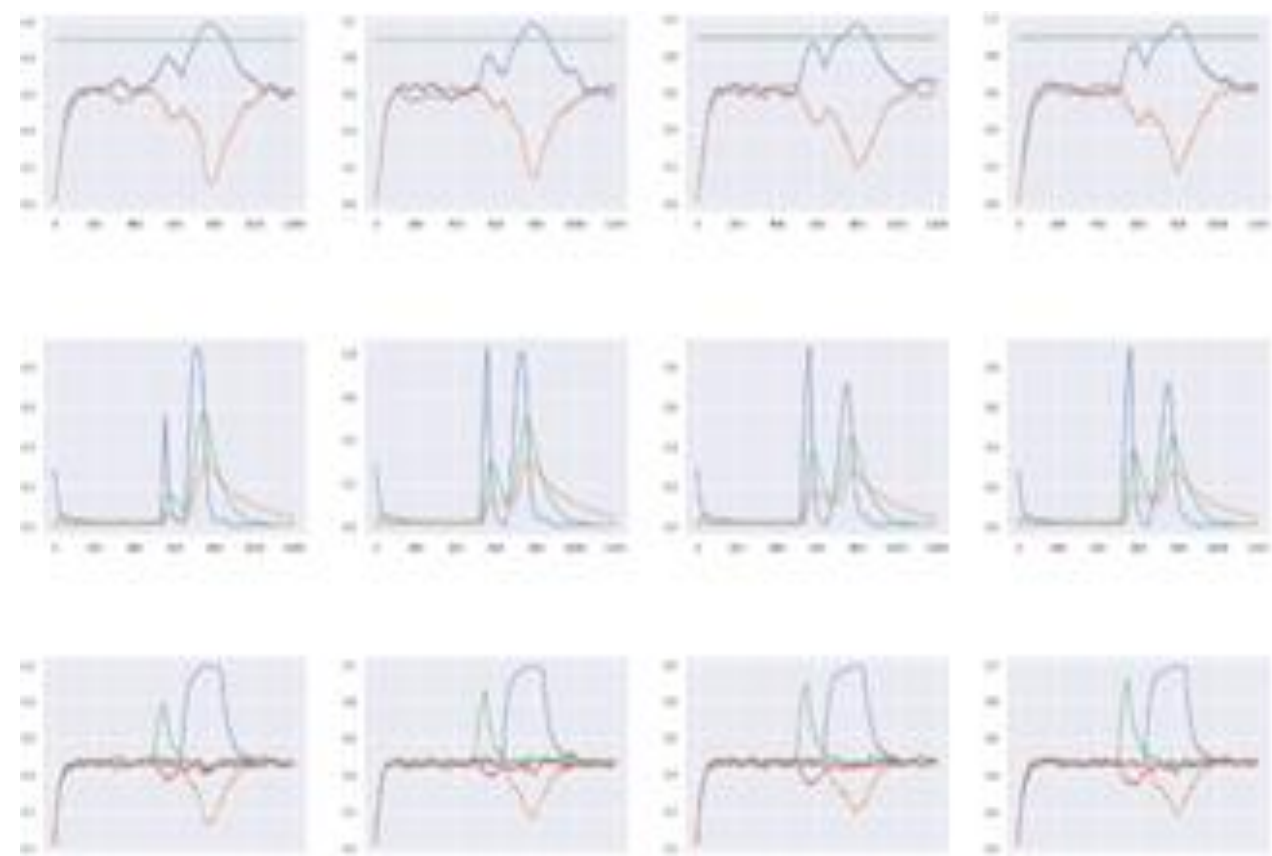

Figure 2: Activation of main layers (Top: ML, Middle: AL, and Bottom: RL) in a single congruent trial where a target (font color with strength 3.0) is preceded by words with different levels of meaning or strength (1.5, 2.0, 2.5, and 3.0), from left to right, respectively. The congruency is clear in the top row, the ML.

While the first stimulus or S1 could be color names (e.g., RED) or non-color words (e.g., CAR) the S2 is supposed to be the font color (the target). In previous models of Stroop (e.g., Servan-Shreiber et. Al., 1990) words have been represented as having a stronger representation compared to colors due to extensive exposure to words in the educational system and other aspects of modern life. Here, for the sake of simplicity, we fixed the connection for the S2 at 3.0 and a longer duration (200 msec) compared to S1 (50 msec) because in the Stroop participants are usually supposed to decide on font colors and ignore the words. This sort of longer S2 compared to S1 is also used in priming models because targets remain longer than the prime (e.g., Sohrabi and West, 2009).

Moreover, there was a short Inter-Stimulus Interval (ISI) between the initialization of RL1 and RL2, 85 msec, because in practice only one stimulus arrives at a time, depending on factors such as perceptual mode, strength, or attention (see also, Anderson, 2007). In the brain, it takes 55-90 msec for LC (and other areas like basal ganglia) when coupled with representational or processing layers (e.g., Usher et al., 1999). Also, as will be noted in the discussion, stimuli features are prioritized during cognitive processing (e.g., Bartsch et al., 2021).

## A Typical Trial

In a typical trial, the two units in the IL1 are clamped to binary values (1 and 0, for left and right units respectively), quickly followed by the IL2 being clamped in the same way, but swapped for incongruent trials (0 and 1, by flipping the units' value). These binary codes represent the two stimuli features, i.e., two color name words (prime) in IL1 and two colors in IL2 (target). These activations spread through connections with different levels of strengths or weights between IL and RL and RL and ML. The weights were fixed at 3.0, except otherwise mentioned, e.g., when the main parameter changes, those between IL1 and RL1 were decreased to represent weak or neutral stimuli, acting like a prime. When any one of the two units in RL1 (S1 or prime) reaches a specified threshold of .75 (Figure 2-3, Bottom, threshold not shown), the AL is triggered with the phasic mode (Figure 2-3, Middle), which in turn causes the neuromodulation of RL and ML (the product of activations and a constant, here 3.5). This is followed by S2 (target) activation and again in turn creates another AL phasic response and the resulting neuromodulations. But both phasic responses can affect each other dynamically. The ML layer (Figure 2-3, Top) is influenced by RL and AL, leading to a response whenever activation of any of its units (left or right) crosses a threshold, here .9. The number of cycles from the presentation of the prime to the emergence of response is considered as Reaction Times (RTs). However, the 500 cycles/msec can be added to RTs to make it more realistic. This is also related to the time that lets the initial activation reach equilibrium and here to make the RTs more realistic because the extra sensory-motor processing time is included in human behavioral data. Other main parameters remained fixed as in the original model (e.g., Sohrabi and West, 2009; also see the companion Github repo).

## Results

We ran a series of simulations, aimed at showing the effect of a neutral stimulus as primes (S1, representing non-color words) on the processing of a target stimulus (S2, i.e., color). As mentioned, this was simulated by decreasing the strength of the connection weights from the input layer to the representation layer (IL-RL, 3.0 for very strong, 2.5 for strong, 2.0 for medium, and 1.5 for weak). As shown in Figure 4 and Table 1, the weaker the prime, the faster was the decision on the target, in both congruent and incongruent trials.

Therefore, the RTs were faster in trials with weaker, compared to trials with stronger S1, resembling the neutral and regular trials, respectively, as happens in RFE. Decreasing this further (close to 0) caused no major differences and it does not make sense as the congruent or incongruent trials both become neutral.

Regarding the difference between congruent and incongruent trials, the well-known congruency effect (i.e., interference) was also simulated without changing any parameters, i.e., better performance in the congruent compared to incongruent trials. These results were attained without including conflict monitoring modules, apparently necessary for paradoxical Negative Congruency Effect (NCE) in priming (Sohrabi and West, 2009), which occurs in the incongruent trials driven by conflict due to the incongruencies between the two stimuli, not happening in neutral and congruent trials.

To further explore the role of attentional resources in RFE, we looked at the differences in the activation dynamics of RL, AL, and ML for the four levels of strength (see Figure 5, for congruent trials, and Figure 6, for incongruent trials). The figures show the average activation in different layers' units for the 20000 runs for the four levels of prime strengths (from neutral or weak to strong, from left to right of each row). As can be seen from left to right, the targets primed by weak stimuli were processed faster and crossed the threshold earlier, both in RL (Bottom row) and ML (Top row) layers.

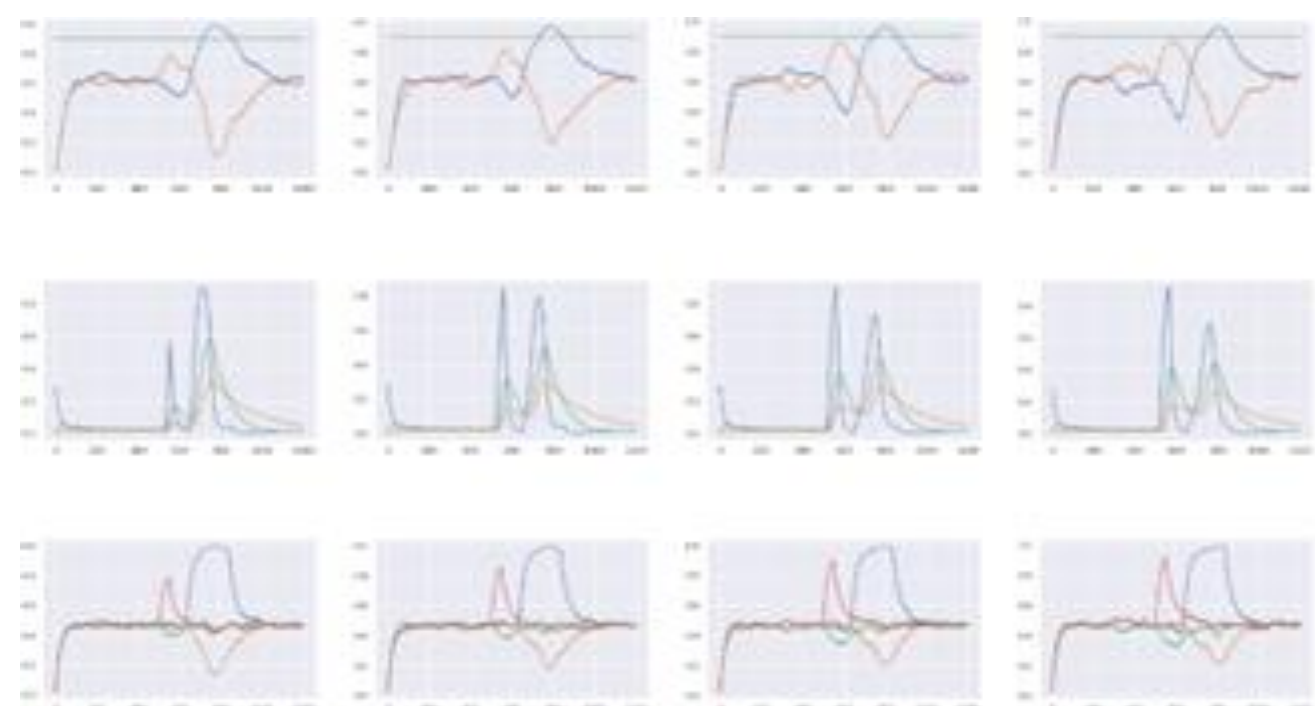

Figure 3: Average activation of main layers (Top: ML, Middle: AL, and Bottom: RL) in a single incongruent trial where a target (font color with strength 3.0) is preceded by words with different levels of meaning or strength (1.5, 2.0, 2.5, and 3.0), from left to right, respectively. The incongruency is clear in the top row, the ML.

As mentioned previously, when a unit in the ML layers reaches a threshold for weaker compared to stronger primes, it leads to the RFE. Similarly, as can be seen from left to right of the figures (Middle row), the weak primes left higher activations in the AL layer for the target, which in turn modulated the RL and ML, causing attentional boosting.

Although, one can clearly see the activation differences for the strength levels in AL from left to right of the row, the results were confirmed by pick prominences measurement of AL activation (for details see the python notebooks in the Github repo as follows). It revealed the activation difference where the attention layer's response to RL for a weak S1 (i.e., non-color words) was less phasic compared to its response to a strong S1 (color words), leaving higher modulatory resources for S2 (Stimulus 2 or target, here colors).

On the other hand, also a strong S1 can highly activate AL, creating high modulatory effects in RL and RL. This may cause premature responses in the form of short RTs, and if the response is not triggered, a decrease or refraction in attentional resources for S2 can result in very long RTs or even misses, i.e., no responses are triggered at all. The model with current parameters and stimuli strength (1.5 to 3.0) caused no premature responses but made some misses, when S1 was strong, mainly due to the refractory period of AL for processing the S2. This is the reason for higher SDs and no further increase in RT for stronger S1 (from 2.5 to 3.0), because misses were removed from the final mean RTs (as shown in Table 1 and Figure 4).

Additionally, we compared the difference between the strength levels in attentional response of AL, a similarity measure known as Fréchet District was employed (Jeckel et al., 2018). Indeed, the difference between lowest to highest values (1.5 to 3.0) was the maximum, and something between for other levels. This was also demonstrated using a joint distribution of the AL activation at the four strength levels. Again, the strength levels more deviant from each other (e.g., 1.5 and 3.0) entailed more complicated and larger differences and became similar for both congruent and incongruent conditions with the high epoch numbers (for details see the python notebooks in the Github reop as follows).

Table 1: Mean and SD for congruent and incongruent conditions at different strength levels, weak (1.5), medium (2.0), strong (2.5), and very strong (3.0). The trials with weak primes (neutral) were processed faster than those with stronger primes, as happens in RFE.

| | Congruent | | Incongruent | |
|---|---|---|---|---|
| Strength | Mean | SD | Mean | SD |
| 1.5 | 209 | 2.71 | 214 | 2.5 |
| 2.0 | 221 | 3.94 | 228 | 3.48 |
| 2.5 | 229 | 7.96 | 240 | 4.45 |
| 3.0 | 230 | 18.6 | 247 | 17.3 |

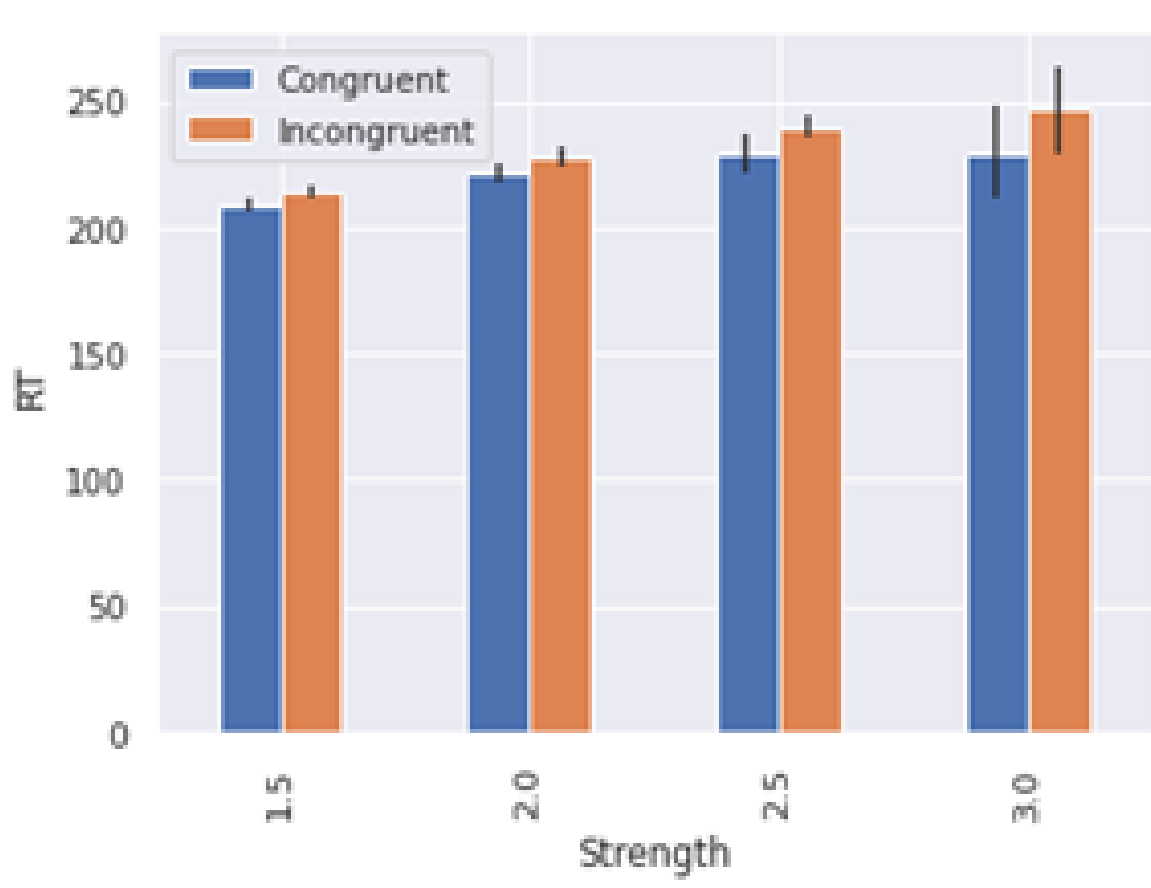

Figure 4: The main simulation result shows the effect of representation strength of Stimulus 1, by changing the strength of the connection weights between the input layer and representation layer (IL-RL), i.e., weak (1.5), medium (2.0), strong (2.5), and very strong (3.0). The trials with weak primes (neutral) were processed faster than those with stronger primes, as happens in RFE.

Moreover, we also found similar results by changing input activation instead of connections as another way to simulate weak or neutral stimuli, simply by decreasing the input activation from 1 (for strong stimulus) to .75 (medium strength stimulus) and .5 (weak stimulus). Again, a further decrease caused no major differences. There are other ways to manipulate the strength of representation (Sohrabi and West, 2009), but they are all reflected in the amplitude of activation functions.

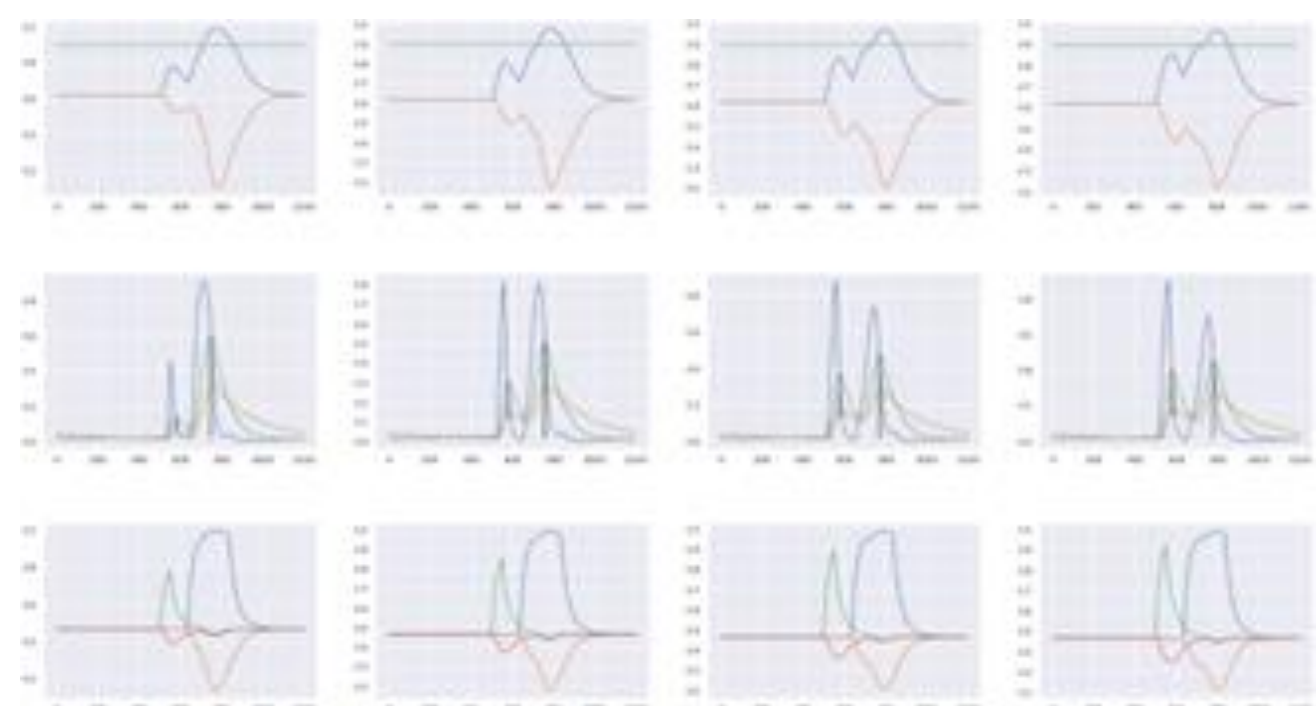

Figure 5: Average activations of all layers in the congruent condition (from top to bottom, ML, AL, and RL, respectively) at four levels of prime strength (1.5, 2.0, 2.5, and 3.0), shown from left to right of each row, respectively. The congruency is clear in the top row, the ML.

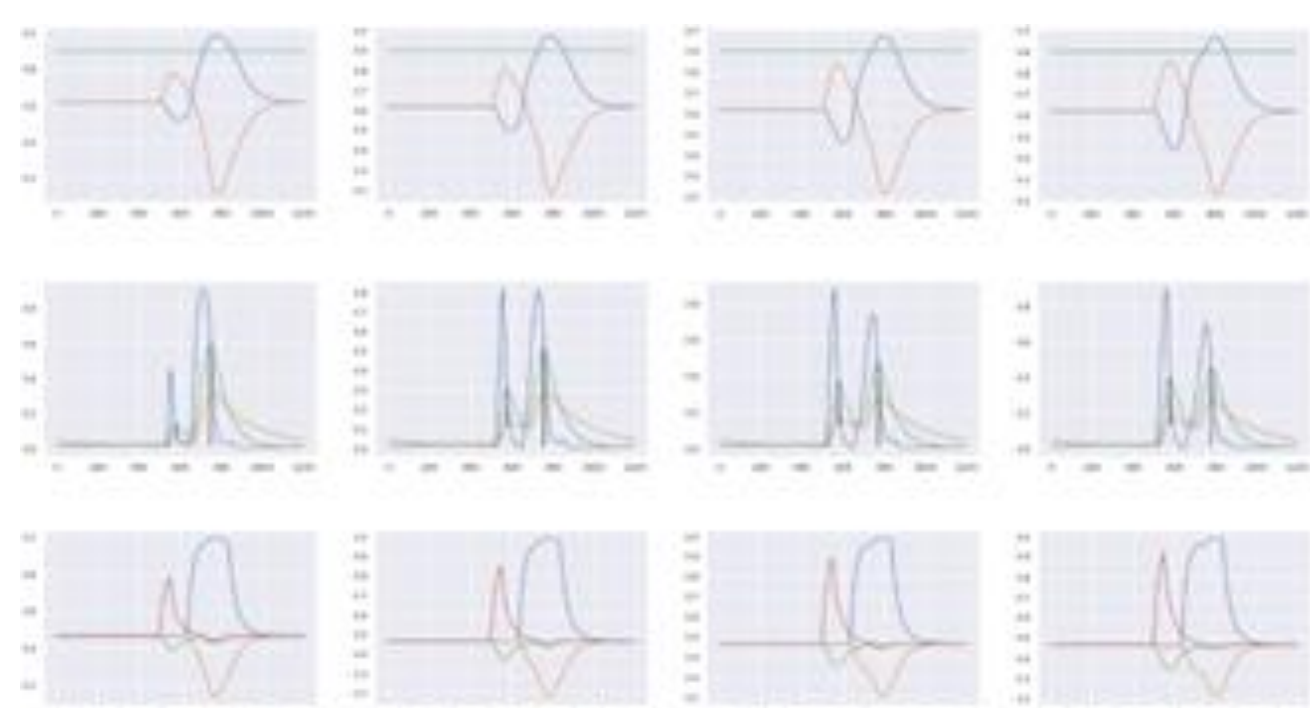

Figure 6: Average activations of all layers in the incongruent condition (from top to bottom, ML, AL, and RL, respectively) at four levels of prime strength (1.5, 2.0, 2.5, and 3.0), shown from left to right of each row, respectively. The incongruency is clear in the top row, the ML.

## Discussion

We showed the ability of a simple neurocognitive model, consisting of perceptomotor and neuromodulatory layers, in simulating a paradoxical phenomenon in the Stroop task, namely the Reverse Facilitation Effect (RFE). The simulations revealed a decreased speed in terms of reaction times for the target stimulus (color names) when paired with another stimulus that has a weak representation, comparatively (i.e., neutral or non-color names in the typical Stroop task). Therefore, there was an increase in RTs for congruent trials when the strength decreased, explaining the slower RTs for the congruent (and incongruent) compared to the neutral condition, i.e., RFE.

In the simulations, the RFE arises from weak attentional activation for neutral stimuli (i.e., non-color or nonword strings), compared to color names, which results in reduced attentional refraction for the target stimuli (font colors). However, when meaningful primes are used (here ones with stronger connections), the refractory period of attention occurs, making the trials slower than those with neutral primes.

Our main finding indicates that the RFE can be modelled without including conflict (in contrast to an existing model, Kalanthroff et al., 2018) but only through the fluctuation in attention, in a dynamic way. This includes the attentional phasic response to the two stimuli as if they are competing on the limited resources, in the so-called System 1. We showed this phenomenon using our previous model of priming effects (Sohrabi and West, 2009) without any special changes to its parameters and yet by making it simpler, i.e., by removing the conflict module, which seems necessary for the Negative Priming Effect (NCE) in priming (may be more involved in System 2).

Based on the simulations, there was an increase in RTs when the prime strength increased, explaining the slower RTs for the congruent condition compared to the neutral (i.e., RFE) as well as the incongruent conditions. Like human data (e.g., Goldfraf and Henik, 2007; see also, e.g., Kalanthroff et al., 2018) the effect turned to RFE. Therefore, the classic Stroop task can be imagined as a priming process (i.e., online priming of font colors by color words). The RFE and other alterations in the congruency effect such as NCE seem to be mainly governed by the refractory period of attention (and at least the former turned out not to be heavily affected by conflict). This reveals a similar basis for this broad range of phenomena. According to the model, attention has two phases, with short Stimulus Onset Asynchronies (SOAs), the processing of the target is facilitated by attention, but with long SOAs it is diminished due to an attentional refractory period. This refractory period is caused by neuromodulation, i.e., a non-linear multiplication, rather than addition or incrimination.

However, in the priming paradigm, the delay between the two stimuli (ISI) is exposed externally by the experimental setup, but it is internal in the Stroop, related to faster processing of words compared to color naming. Also, it seems that in both tasks one stimulus is usually weaker. In the priming, S1 comes early and is usually masked, but in the Stroop, the S1 and S2 appear at once. When the S1 triggers a low attentional response, it causes a smaller refractory period for the S2. Here we suppose that words (even if neutral) have a stronger representation compared to font color, so it triggers attention before S2. The exact prioritization and temporal process require more investigations. While the two stimuli can share the same attentional resource, usually they are not processed at once, ending up in a competitive race, depending on their strength of representation and ISI (Sohrabi and West, 2008; 2009). Changing the strength of the two stimuli can dynamically affect the direction of Stroop or priming effects. A strong stimulus causes a refractory period of attention for the other one, something like visual silencing (e.g., Kok et al., 2011). Although Kok and colleagues have attributed visual silencing to early visual processing, not attention, here we do not explicitly separate their joint effects. When a stimulus is neutral or weak, the phasic attention period is allowed to evolve in response to the other stimulus. This causes better performance in the neutral trials compared to congruent as if attentional modulation is dedicated to one stimulus and is not shared between the two, causing the RFE.

Now we know that some types of attention are more stable than others (Snyder et al., 2021) and neutral stimuli differ in their levels of meaningfulness (Hershman et al., 2021). So, the more meaningful ones activate more resources, e.g., BED is more like RED when compared with BIT. These different stimulus types and/or their different stimuli features, including distractors, are prioritized for further processing (Bartsch et al., 2021). So, to uncover the main underlying process in RFE, we focused on intra-trial attention (through a single trial) that presumably causes minimum conflict, compared to planned and context-dependent tasks (Ott et al., 2021). As Snyder and colleagues showed, this type of attention seems to involve fewer variables and fewer higher-order processes, mainly in the occipital cortex, far from the prefrontal cortex.

On the other side, the Kalanthroff and colleagues' simulation of RFE (Kalanthroff et al., 2018) was based on Kornblum and colleagues (Kornblum, 1994; Kornblum and Lee, 1995) notion of different kinds of conflict. Among them, their Type 8, involved in incongruency effect, is related to what they called coding and response conflict as well as information conflict, the inconsistency between or within different stimuli or responses. Our model has within-layer lateral inhibition which can be thought of as this type of conflict. But usually, conflict is imagined as being at a higher level than that, for recruiting monitoring and control processes (e.g., Servan-Schreiber et al., 1990; McClure et al., 2015). Another type of conflict, namely task conflict, employed by Kalanthroff et al. (2018) to simulate RFE, is supposed to happen at a higher level, i.e., coding level. Similar views and models support this dynamic and multi-level view. They all accept that attentional control depends on many factors (Ritz et al., 2021) and that the best way to model it might be through a hierarchical view (Verbeke and Verguts, 2020; Markovic et al., 2021). A more complicated and improved model might be needed especially for between-trial conflict and strategic control of attention.

Therefore, for the big picture, this simple model might not be enough as a full explanation of all sorts of effects, even for RFE. Thereby, regarding the simplicity of our model compared to others', we must see its behavior when considering more complex processes. Another model that has similarly employed attention resources to simulate Stroop effects including RFE is the one by Caron and Stewart (2020), though they call it negative facilitation effect. Their spiky neurons model relates RFE to allocating more attention to the color target not by withdrawing attention to the color words in the current model. This is another challenging issue to be addressed in the future to compare our model which is more reduced and abstract to more detailed but less parsimonious neuronal models such as spiky or other types of neural networks.

Moreover, we have already claimed (Sohrabi and West, 2008) that all factors, including stimulus types, durations, and SOA, alter the representational strength and hence change the direction of the congruency priming effect. For generalizing this claim to Stroop, we can conclude that neutral and other low strength stimuli can change the Stroop effects, among them the RFE. As we saw in the simulation results, depending on the strength of the two stimuli, the result is in favor of one of the two, in a dynamic, bifurcation way. The model predicts that other factors that can affect the strength of representation may result in the alteration of congruency effects. For example, words in second languages, emotional words, word frequency and regularity, and even age of acquisition of the words all may affect the Stroop effects, including RFE.

Furthermore, the model in addition to simulating RFE (better performance in neutral compared to congruent) showed the incongruency effect as well (better performance in the congruent compared to incongruent). Therefore, it was not an overfit model to show only the difference between congruent and neutral trials. The model has been previously employed for several priming phenomena (see, Sohrabi and West, 2009) that can help explain RFE even further. For example, in that previous work the RTs were faster in weak mask conditions (like neutral stimuli in Stroop) than strong mask or strong prime conditions. Regarding the color and word competition in Stroop, the weak or neutral stimuli might reach the threshold (conscious or unconscious) slowly. This internal SOA or the so-called "endogenous or semi-SOA" (Sohrabi, 2008) leads to congruency reversal in priming, positive to negative, and presumably the Stroop effect to RFE.

However, still, it is not clear whether this model has other advantages over Kalanthroff et al.'s (2018), especially without the conflict module as was used here, something to remain for future studies. This might bring more insights into this domain as some authors have raised questions about some aspects of our model (Whitmarsh et al., 2013; Gaschler et al., 2014). Therefore, showing the ability of the model in simulating wider cognitive phenomena is worth considering in later improvements.

Finally, while the cognitive science continues toward a unified model, the current study can add more intuition to the previous similar dynamic models of Stroop (Kalanthroff et al., 2018; see also Caron and Stewart, 2020), Priming (Sohrabi and West, 2009), and attentional blink (Nieuwenhuis et al., 2005). But a big picture of the underlying mechanism is hidden, so incorporating both context and time (Hitchcock et al., 2022) can bring more explanations and applications. Especially, the dynamic fluctuation in attention and the resulting fragility, demonstrated through different ways, might be used as a measure of congruency effect for different types of speed and accuracy scales. Any behavioral/cognitive index for this, as a correlate of neurological or psychological pathology might be helpful, as well as improvements to the current interventions such as cognitive bias modification and other attentional and cognitive training tools.

**Acknowledgement**

An early version of this work was presented as a poster at the International Conference on Brain Engineering and Neuroscience, Tehran, Iran, 2018.

**Code Availability**

The code is available at the following Github repo: https://github.com/ASohraB/PaperRFE